\documentstyle[12pt]{article}   
\newcommand{\partd}[2]{\frac{\partial #1}{\partial #2}}
\newcommand{\bc}{\begin{center}} 
\newcommand{\ec}{\end{center}}        
\newcommand{\be}{\begin{equation}} 
\newcommand{\ee}{\end{equation}}        
\newcommand{\ba}{\begin{eqnarray}} 
\newcommand{\ea}{\end{eqnarray}}        
\newcommand{\bi}{\begin{itemize}} 
\newcommand{\ei}{\end{itemize}}        
\begin{document}
\begin{center}  
{\Large\bf  
  Metric structures of inviscid flows}  
\end{center}  
\begin{center}  
{\large Rub\'en A.\ Pasmanter}\\   
{\em   
KNMI, P.O.Box 201, 3730 AE \quad De Bilt\\  
e-mail: pasmante@knmi.nl}   
\end{center}  

 \begin{center}   
 {\Large Abstract}   
 \end{center}   
 \begin{center}   
 {\begin{minipage}{4.5in}   
 \small   
 An intrinsic metric tensor, a flat   
 connexion and the corresponding  
 distance-like function are constructed in the  
 configuration space formed by the velocity field  
 {\bf and} the thermodynamic variables of an inviscid   
 fluid.   
 The kinetic-energy norm is obtained as a limiting case;
 all physical quantities are Galilean invariant.
 Explicit expressions are given for the case of an ideal 
 gas. 
 The flat connexion is {\bf not} metric-compatible. 
 These results are achieved by applying  
 the formalism of statistical manifolds  
 \cite{amari,otros} to the statistical mechanics of a  
 moving fluid. 
  \end{minipage}}  
 \end{center}
 \begin{center} 
  Submitted for publication. 
 \end{center}  
 \section{Introduction} 
 \vspace*{-0.1cm} 
 Typically, 
 the configuration of a simple fluid 
 at a certain time 
 $t$ is described by its density field 
 $\rho(\vec x,t)$, its temperature field 
 $T(\vec x,t)$ and its velocity field 
 $\vec u (\vec x,t)$ where $\vec x$ is the 
 position vector in a $d$-dimensional space.
 This $d$-dimensional space is often the 
 everyday 
 three-dimensional Euclidean space or a 
 two-dimensional surface embedded 
 in three-dimensional
 space, e.g., a plane or a sphere. 
 In that $d$-dimensional space, it is possible to 
 calculate the square of the distance 
 between two nearby positions, 
 say $\vec x$ and 
 $\vec x + d\vec x$, in the form of an expression 
 \be  
 M_{\lambda\nu}(\vec x) dx^\lambda dx^\nu 
 \label{rie} 
 \ee 
 where $M_{\lambda\nu}(\vec x)$ is the 
 metric tensor, $dx^\lambda$ is the 
 $\lambda$-component of $d\vec x$ in some coordinate 
 system, $1 \le \lambda \mbox{\ and\ } \nu \le d$, 
 and the convention of 
 summation over repeated lower and upper 
 indices holds. 
 For example, on a two-dimensional 
 ($d=2$) sphere of radius $r$, 
 in the coordinate system defined by the angles $\vartheta$ 
 and $\phi$ (latitude and longitude), 
 the non-vanishing components of the metric tensor are
 \ba 
 M_{\vartheta\vartheta} = r^2,\\ 
 M_{\phi\phi} = r^2 \sin^2\vartheta, 
 \ea 
 and the corresponding volume element is 
 $\sqrt{\det\,(M_{\lambda\nu})}= r^2\sin\vartheta$. 
 Such a metric tensor is also what 
 is required in order to compute the scalar 
 quantity we call the norm  
 of a (tangent) vector, e.g., the norm 
 of the velocity vector
 $d\vec x/dt$ is given by 
 \be 
 M_{\lambda\nu}(\vec x) \frac{dx^\lambda}{dt\,} 
 \frac{dx^\nu}{dt\,} .
 \label{velo} 
 \ee
 Similarly, one can compute the angle between two
 (tangent) vectors, e.g., two velocities, at  
 each position in the $\vec x$-space. 

 The situation in the configuration space of a 
 fluid with coordinates $(\rho , T, \vec u)$ is 
 totally different from that in the $\vec x$-spaces 
 described above as long as it is not known how to 
 introduce, in a natural and intrinsic way, a geometric 
 structure, e.g., something similar to the metric tensor 
 $M_{\lambda\nu}(\vec x)$ 
 in (\ref{rie}) and (\ref{velo}). 
 Such a lack of geometric    
 structures  imposes many restrictions on the   
 kind  of computations that one can perform:    
 it is impossible to talk of ``the distance" between   
 two states of the fluid, i.e.,   
 between two positions in the space $(\rho , T, \vec u)$;   
 neither is it possible to talk of ``the norm"    
  of the vector formed by the rate of   
 change of the dynamical variables $(d\rho/dt , dT/dt,
 d\vec u/dt)$;   
 it is not possible   
 to consider the angle between two such vectors;   
 there is no volume element defined in configuration space, 
 therefore, it does not make sense to talk about ``the  
 density" of a distribution of points in that space;   
 etc.  
 These limitations are too restrictive since, 
 to name but a few examples: 1) A norm is needed when 
 studying the (in)stability of flows, especially when the 
 linear growth of  
 the perturbations is only transient \cite{energynorm};\,  
 2) The angle between two directions is required in order 
 to determine how strongly a given perturbation projects 
 along an optimal perturbation \cite{Farrell};\, 
 3) A volume element is needed in order to determine the 
 density distribution of ensemble simulations of flows 
 \cite{Mureau}.  
 Therefore, it is often opted to   
 circumvent these limitations by introducing an acceptable,  
 if somewhat arbitrary, metric  
 tensor, e.g., the kinetic-energy metric in the case  
 of incompressible, isentropic flows \cite{energynorm}. 
 Such a situation is 
 not unique to fluid mechanics: often   
 our knowledge of a dynamical system is   
 limited to the variables that define its   
 phase space and to the equations that   
 determine the time evolution of these  
 variables  while any   
 information on the geometry of the 
 phase space is lacking.

 In this article it is shown how to generate,  
 in a natural way,   
 not only the analogue of the spatial metric 
 tensor $M_{\lambda\nu}(\vec x)$ but also  
 a rich geometric structure  
 in the configuration space of 
 an inviscid, moving fluid starting from the fluid's  
 probability distribution. 
 The derivation of these results makes use 
 of ideas and techniques from statistical mechanics, 
 see e.g., \cite{balescu,grand} and from statistical 
 manifolds theory, see e.g., \cite{amari,otros}, a 
 particular branch of differential geometry, see e.g., 
 \cite{geometry,Erwin}. 
 The presentation is as self-contained as possible.  
 A complete description of the required background material,   
 however, was not attempted; 
 detailed bibliographic references are given for the 
 benefit of those readers willing to go deeper into 
 some aspects.\\  
 The paper is structured as follows: 
 In Section 2, the probability 
 density of a moving fluid in local thermal equilibrium 
 is presented. 
 In Section 3, some 
 fundamental concepts of
 the theory of statistical manifolds are introduced.   
 In Section 4  
 these ideas are applied to a moving fluid, the 
 natural metric tensor in the fluid's configuration 
 space is derived and an application is 
 described; 
 some additional information is given in  
 Appendix A. 
 In Section 5, it is shown that the probability 
 density generates a flat, but non-Riemannian,
 geometry in the configuration space. 
 Basic concepts on flatness (of affine connexions) 
 are reviewed in Appendix B.
 Based on this flatness, a distance-like function is  
 defined in Section 6. 
 The computed metric and distance-like function are 
 Galilean invariant, as they should. 
 When the thermodynamic variables can be ignored, 
 both the metric and the distance-like function 
 reduce to the above-mentioned  
 kinetic-energy norm times a conformal factor. 
 For the purpose of illustration, the expressions 
 corresponding to an ideal gas are explicitly given in 
 Sections 4 and 6.  
 In Section 7, we comment on related  
 work done by others; in Appendix C,    
 the differences between the approach
 developed in the present paper and a particular 
 Hamiltonian formulation of fluid dynamics are 
 pointed out.
 Finally, in Section 8, we review the results,  
 discuss their generalizations, e.g., to systems far 
 from thermal equilibrium. 
 \section{Statistical mechanics of a moving fluid} 
 \vspace*{-0.1cm} 
 The first step in our construction of an intrinsic 
 geometric structure in the $(\rho , T, \vec u)$ 
 space consists in associating a probability density 
 to these variables. 
 In this Section we 
 derive the fluid's probability distribution
 by applying statistical mechanics to 
 a moving fluid in local thermal equilibrium. 
 
 Consider a simple fluid characterized, on a       
 macroscopic level,       
 by its density field $\rho(\vec x,t)$, temperature        
 field $T(\vec x,t)$ and velocity field        
 $\vec u(\vec x,t)$; 
 from here onwards, we do not express the
 $(\vec x,t)$-dependence explicitly.
 We adopt the standard assumptions that make        
 possible the derivation of the Euler equations     
 and, less rigorously, of the 
 Navier-Stokes equations from local thermal        
 equilibrium statistical mechanics, 
 see, e.g., \cite[Chapters 1 to 3 of Part 1]{Spohn} and  
 \cite[Chapters 1 to 3]{Demasi}, to wit:  
 One considers systems in which the 
 above-mentioned fields vary on length scales 
 that are much larger than the intermolecular 
 distances and on time scales that are much larger
 than the microscopic time scales  
 needed for local thermal equilibration. 
 Then the space can be divided 
 into small volume elements 
 $\Delta V$ with a typical size much smaller than 
 the length scales of the macroscopic fields but 
 much larger than the intermolecular distances so 
 that the 
 statistical mechanical description applies to them,
 i.e., on the length scales  
 of $\Delta V$, the fluid is in local thermal equilibrium. 
 The dynamics of the fluid is assumed to be 
 such that the following
 extensive quantities are conserved: 
 a) 
 the number
 of (indistinguishable) particles $N$,    
 b) 
 the total momentum of these particles   
 \be    
 \vec M :=m\sum_k^N\vec v_k, 
 \label{M}  
 \ee  
 where $m$ is the mass of the particles,
 $\vec v_ k$ is the velocity of the $k$-th particle,  
 and 
 c) their total energy    
 \be    
 E:=\sum_k^N{\left( \frac{1}{2}m   
 |\vec v_k|^2 +   
 \frac{1}{2}\sum_{l\neq k}^N V_{kl}\right) }, 
 \label{E}   
 \ee    
 $V_{kl}$ being the potential interaction between   
 particles $k$ and $l$. (The symbol to the left 
 of $:=$ is defined by the expression to the right.)

 Let us denote the conserved quantities by  
 $\{ H_i (\xi_N)\, |i=1,\dots ,s\}$ where 
 $\xi_N := \{\vec x_1,\dots ,\\ 
 \vec x_N ,  m\vec v_1 ,\dots 
 ,m\vec v_N\, | N=1, \dots , \infty\}$ 
 and the values of $\{ H_i\}$ in a 
 thermal equilibrium state by $\{\eta_i \}$. 
 These two elements, i.e., the expressions defining 
 the conserved quantities $\{ H_i \}$ as functions of the 
 microscopic variables $\{ \xi_N \}$ and the values 
 $\{\eta_i \}$ characterizing the macroscopic state, are the 
 essential ingredients needed for the statistical mechanical 
 description of the  system.
 Statistical mechanics establishes that 
 the probability density in the phase space 
 $\xi_N $,  
 of a system in $\Delta V$, in thermal equilibrium with and  
 free to exchange particles, momentum and energy with a 
 surrounding thermal bath is  
 \be 
 p(\xi_N; \theta) := 
 \exp {\left[ \theta^i H_i(\xi_N) \right] } 
 /c^N N! {\cal Z}(\theta), 
 \label{pd} 
  \ee 
 see, e.g., \cite[Chapter 4]{balescu}, 
 \cite[Chapters 9, Section B.3]{grand}. 
 We use the standard short-hand conventions: 
 repeated lower and upper indices are 
 summed up; $\theta$ stands for $\{\theta^i|i=1, 
 \dots , s\}$.  
 The ${\cal Z}(\theta)$ in the denominator, 
 called the grand partition function, 
 is the normalizing factor needed to make 
 sure that 
 $ 
 \sum_N^\infty 
 \int\! d\xi_N \, p(\xi_N; \theta) =1,   
 $ 
 i.e., 
 \be 
 {\cal Z}(\theta) := 
 \sum_N^\infty 
 \int\! d\xi_N \,
 \exp {\left[ \theta^i H_i(\xi_N) \right] } 
 /c^N N! ,
 \label{Z} 
 \ee 
 the constant $c$ makes each of the terms 
 contributing to ${\cal Z}$ dimensionless. 
 From (\ref{pd}) and  (\ref{Z}) we see that the
 average values of the conserved quantities 
 $\{H_i(\xi_N)\}$, which we call $\{\eta_i\}$, 
 can be written as 
 \be 
 \eta_i (\theta) = 
 \frac{\partial \ln {\cal Z}(\theta)}%
 {\partial{\theta^i}},\quad i=1,\dots ,s . 
 \label{eta} 
 \ee 
 Since the thermal equilibrium state of a system with  
 $s$ conserved quantities is completely 
 characterized by their macroscopic values 
 $\{\eta_i\}$, 
 the last expression makes evident that the whole 
 thermodynamics can be obtained from the grand 
 partition function ${\cal Z}(\theta)$. 
 In this way, one identifies $\ln {\cal Z}(\theta)$ 
 as the thermodynamic potential which is minimized 
 at fixed values of the  
 $\theta$-variables and the $\theta$-variables as the 
 intensive thermodynamic parameters (like temperature, 
 pressure, chemical potential, etc).  
 Since the system in $\Delta V$ is in thermal 
 equilibrium with its surroundings, the values of 
 these intensive variables must be 
 equal to those of its thermal bath,
 see, e.g., \cite[Chapter 4]{balescu}, 
 \cite[Chapters 9, Section B.3]{grand}.

 In the case of a moving fluid,  
 the conserved quantities are  
 $(N,E,\vec M)$, confer (\ref{M}) and (\ref{E}), 
 so that $s= d + 2$ and      
 the probability density (\ref{pd}) reads         
 \be       
 p(N,E,\vec M;\gamma,-\beta,\vec\kappa) =        
 \exp [\gamma N -\beta E + \vec\kappa\cdot\vec M ] /        
  c^N N! \cal Z(\gamma,-\beta,\vec\kappa),        
  \label{probability}        
 \ee 
 i.e., $\theta = (\gamma,-\beta ,\vec \kappa)$. 
 It follows then that the average, macroscopic 
 values of $(N,E,\vec M)$ are given by
  \ba  
 \langle N\rangle =   
 \partd{\ln{\cal Z}(\gamma,-\beta,\vec\kappa)}{\gamma}
 \label{relation1}\\  
  \langle E\rangle =  
 -\partd{\ln{\cal Z}(\gamma,-\beta,\vec\kappa)}{\beta}
 \label{relation2}\\  
 \langle M_\lambda \rangle=
  \partd{\ln{\cal Z}(\gamma,-\beta,\vec\kappa)}{\kappa^%
 \lambda},\quad\lambda =1,\dots , d,   
 \label{relation3}  
 \ea   
 where the pointed brackets indicate average over  
 the probability distribution (\ref{probability}). 
 It is convenient to introduce the macroscopic velocity 
 of the fluid $\vec u$ by
 \be 
  m \langle N\rangle \vec u := \langle\vec M\rangle 
 \label{u}
 \ee 
 and the specific internal energy 
 $\epsilon(\gamma,\beta,\vec\kappa)$ as
 \be 
 \langle N\rangle ( \epsilon + \frac{m}{2} u^2 ) := 
 \langle E\rangle .
 \label{epsilon} 
 \ee 
 Relating these expressions to the thermodynamics of 
 the system, one identifies $\ln {\cal Z}(\gamma, 
 -\beta,\vec \kappa)$ as the  
 thermodynamic grand potential, i.e., 
 \be  
 \ln {\cal Z}(\gamma,-\beta,\vec\kappa)=  
  \beta \Delta V  P,  
 \ee  
 where $P= P(\gamma,\beta,\vec\kappa )$ 
 is the pressure,    
 $\beta$ as the inverse local temperature,  
 $\beta\equiv (kT)^{-1}$, $k$ is Boltzmann's 
 constant and $\gamma$ as 
 $\gamma =  \beta\mu$ 
 with $\mu$ the local chemical potential; 
 see, e.g., \cite[Chapter 4]{balescu}, 
 \cite[Chapters 9, Section B.3]{grand}.
 Due to the simple dependence of $\vec M$ and of 
 $E$ upon the particles' velocities $\vec v_k$, 
 confer (\ref{M}) and (\ref{E}), the integrals over 
 the velocities in (\ref{Z}) can be performed; 
 from (\ref{relation3}) and (\ref{u}) 
 one obtains then that 
 \be 
 \vec\kappa= \beta\vec u 
 \label{kappa} 
 \ee 
 and that 
 \be        
 {\cal Z}(\gamma,-\beta,\vec\kappa)=       
 {\cal Z}(\gamma - \frac{m\kappa^2}{2\beta},-\beta,\vec 0).        
 \label{shift} 
 \ee        
 Galilean invariance implies then that 
 \be  
 \gamma = \bar{\gamma} + \frac{m\kappa^2}{2\beta},  
 \label{gamma}  
 \ee  
 where $\bar{\gamma}$ is the value of $\gamma$ for the same  
 system at rest.   
 Making use of (\ref{kappa}), one sees then that 
 $\mu = \bar\mu + m u^2/2$, i.e., the  
 chemical potential of the fluid moving with velocity  
 $\vec u$ is shifted by an amount $m u^2/2$ with  
 respect to $\bar\mu$, the chemical potential of the 
 fluid at rest. 

 In the theory of statistical manifolds \cite{amari,otros} 
 it is shown that a probability density like 
 $p(\xi_N; \theta)$ 
 in (\ref{pd}) induces a geometric structure in the 
 parameter-space $ \theta$, i.e., in the space 
 of the intensive thermodynamic variables. 
 This is worked out in the following Sections.  
 \section{Statistical manifolds} 
 \vspace*{-0.1cm}  
 Given two probability densities, say 
 $p(\xi_N ,\theta_1)$ and $p(\xi_N ,\theta_2)$, 
 how should one quantify their (dis)similarity? Or 
 given three distributions corresponding to $\theta_1 ,
 \theta_2$ and $\theta_3$ respectively, 
 is it possible to determine which pair of distributions 
 is ``closer" than the two other pairs? 
 These and related questions have been extensively 
 studied in statistics, see, e.g., \cite[pages 
 290--296]{enciclo} and \cite{eguchi}. 
 In this Section we give a fleeting overview of   
 some concepts and results from the theory of statistical 
 manifolds that we apply later to a moving fluid.

 If $D(\theta_1 ,\theta_2)$ is a scalar quantity measuring the 
 difference between two probability densities $p(\xi_N,\theta_1)$ 
 and $p(\xi_N,\theta_2)$, then it should satisfy:\\
 1) $D(\theta_1 ,\theta_2)\geq 0$,
 the equality holding if, and only if,   
 $p(\xi_N,\theta_1)\equiv p(\xi_N,\theta_2)$,\\ 
 2) $D(\theta_1 ,\theta_2)$ is sufficiently continuous in 
 $p(\xi_N,\theta_1)$ 
 and $p(\xi_N,\theta_2)$.\\ 
 In the statistical literature such measures are often called 
 divergences or contrast functionals;
 in order to avoid any possible confusion with 
 the usual mathematical meaning of divergence, we shall keep 
 to the second name.\\ 
 We shall impose two additional constraints:\\
 3) The value of $D(\theta_1 ,\theta_2)$ should be independent 
 of our 
 choice of coordinates in $\xi_N$-space, i.e., it should be 
 invariant under general coordinate transformations 
 in the space of the random variables.\\ 
 4)  
 We shall consider only contrast functionals that are local 
 in the random variables $\{ \xi_N\}$\footnote{C.R.\ Rao has 
 considered also non-local contrast functionals, 
 \cite[pp. 226--227]{otros}.}.\\ 
 In order to satisfy these constraints, 
 it is sufficient that 
 the $D(\theta_1 ,\theta_2)$s be 
 of the following form 
 \cite{amari}, \cite[pp.\ 349--350]{eguchi},
 \be 
 D(\theta_1 ,\theta_2) =  
 \int\! d\xi_N\, p(\xi_N,\theta_2)
 f \left( 
 \frac{p(\xi_N,\theta_1)}{p(\xi_N,\theta_2)} 
 \right), 
 \label{D}  
 \ee 
 where 
 the function $f$ must be  sufficiently 
 smooth and 
 should satisfy $f(1)=0$ and 
 $f(t)- f'(1)(t -1) > 0$ if $t\ne 1$.   
 Without any loss of generality, one normalizes $f$ such 
 that $f''(1)=1$.\\

 Consider next two distributions that are infinitesimally 
 close, i.e., $\theta$ and $\theta + d\theta$. 
 It turns out that, up to third 
 order in $d\theta$, the corresponding contrast is 
 given by \cite[theorem 3.10]{amari} 
 \be 
 D(\theta,\theta + d\theta) =  
 \nonumber\\ 
 \frac{1}{2}\,  
  g_{ij}(\theta)\, d\theta^i\, d\theta^j +  
 \frac{1}{2}\left\{ { 
 [i,j;k] + \frac{\alpha}{ 3!} T_{ijk}(\theta) 
 } \right\} 
 \, d\theta^i\, d\theta^j\,  d\theta^k  
 + O(d\theta^4), 
 \label{taylor} 
 \ee  
 where  
 the components of $g_{ij}(\theta)$, 
 known as the Fisher tensor \cite{fisher,Rao}, 
 are given by
  \be       
 g_{ij}(\theta) :=  
 \left\langle {\partd {\ln p}{\theta^i}  
 \partd{\ln p}{\theta^j} }\right\rangle , 
  \label{g} 
 \ee 
 the pointed brackets indicate an average taken over the 
 $p(\xi_N ,\theta)$ distribution.  
 From this expression it is evident that $g_{ij}$ is 
 symmetric in the indices $i \mbox{ and } j$ and that, 
 since it is an addition of products of 
 covariant vectors, under a general 
 coordinates transformation 
 it transforms like the 
 product of two covariant vectors, i.e., 
 it is a covariant tensor\footnote{For a 
 more detailed meaning of these geometric 
 concepts, see, e.g., \cite{geometry,Erwin}.}. 
 Moreover, any sensible choice of the parameters 
 $\theta$ ensures that it is non-singular.\\ 
 The first term in the coefficient of the third-order 
 contribution to the expansion (\ref{taylor}), 
 $[i,j;k]$, is given by 
 \be 
 [i,j;k] := \frac{1}{2} \left( \partd{g_{ik}}{\theta^j} 
 + \partd{g_{jk}}{\theta^i} - \partd{g_{ij}}{\theta^k} 
 \right), 
 \label{christo} 
 \ee 
 its meaning is discussed in Appendix B and in Section 5.  
  The second term in the third-order coefficient is a
 totally symmetric covariant tensor, 
 often called the skewness, 
 \be  
 T_{ijk}(\theta) 
 := \left\langle {\partd{\ln p}{\theta^i}
 \partd{\ln p}{\theta^j}  
 \partd{\ln p}{\theta^k}}\right\rangle.    
 \label{third}  
 \ee  
 Both symmetric tensors, $g_{ij}$ and $T_{ijk}$, 
 will play an important role in the sequel.\\ 
 The constant $\alpha$ in (\ref{taylor}) is given by 
 $\alpha:= 2f'''(1) + 3$. 

 A number of points are worth noticing:\\ 
 $\bullet$\quad 
 The symmetric contravariant tensor $g_{ij}(\theta)$ has all 
 the credentials for being the natural metric tensor in 
 the $\theta$-space. 
 Besides the clear meaning ensuing from (\ref{taylor}), 
 it has an 
 important statistical significance \cite{Rao,Cramer} 
 that is described in Appendix A.\\  
 $\bullet$\quad 
 Up to second order, the expansion (\ref{taylor}) is  
 independent of the particular choice of the function $f$ 
 that defines the contrast $D$, confer (\ref{D}). 
 (Remember that $f$ has been normalized so that $f''(1)=1$.)\\  
 $\bullet$\quad 
 Up to third order, the whole dependence upon the function $f$  
 has been brought back to a single number,
 namely, to $\alpha$.\\ 
 \section{Fisher's metric} 
 \vspace*{-0.1cm} 
 In this and the following Sections, we apply the 
 theory of statistical manifolds to the probability 
 density $p(\xi_N ,\theta)$ of a moving fluid.\\ 
 As shown in the previous Section, 
 the Fisher tensor appears naturally as 
 the metric in $\theta$-space. 
 From (\ref{probability}) it follows that  
 in the coordinate system 
 $(\theta^1,\dots ,\theta^{d+2}) = 
 (\gamma,-\beta,\vec\kappa)$, 
 the metric tensor (\ref{g}) reads,      
 \be      
 g_{ij}(\theta)=  
 \frac{\partial^2\ln {\cal Z}(\theta)}%
 {\partial \theta^i\partial\theta^j}=  
 \Delta V\,\frac{\partial^2\beta P}%
 {\partial\theta^i\partial\theta^j} ,      
 \label{coordina} 
 \ee       
 see also (\ref{apeng}) in Appendix A. 
 We see then that 
 the metric elements are expressed in terms of 
 the derivatives of the thermodynamic potential.  
  Moreover, from (\ref{relation1}--\ref{relation3}) or 
 from (\ref{eta}) one has that, 
 in the $(\gamma,-\beta,\vec\kappa)$ coordinate  
  system,  
 \be  
 d\eta_i = 
   g_{ij}(\theta) d\theta^j, 
 \label{lower}   
 \ee 
 where $\eta_i$ are the average values of the extensive 
 quantities given in (\ref{relation1}--\ref{relation3}). 
 I.e., in this coordinate system, the lowering  
 of the indices corresponds to passing from the intensive 
 variables $\theta$ to the extensive ones $\eta$. 
 The inverse transformation is achieved by raising the 
 indices by means of the inverse metric $g^{ij}(\theta)$. 

  It is instructive to work out the metric tensor for  
 the case of an ideal gas, i.e., for a vanishing 
 intermolecular potential $V_{kl}$ in (\ref{E}).  
 In this case all the integrals over the molecular 
 velocities $\vec v_k$ and over their positions 
 $\vec x_k$ in (\ref{Z}) can be performed and one 
 finds that     
 \be      
 \ln{\cal Z}(\gamma,-\beta,\vec\kappa) = \Delta V \,  
 \beta^{-d/2} \exp (\frac{m\kappa^2}{2\beta} -\gamma).      
 \ee        
 It is convenient to express the results in   
 the coordinate system $(\rho,\beta,\vec u)$, where 
 $\rho$ is   
 the mass density, i.e.,     
 $\rho := m\langle N\rangle/\Delta V$.   
 The non-vanishing elements of the metric are,  
 \ba     
 {g}_{\rho\rho}= \Delta V({m\rho})^{-1}, 
 \nonumber\\  
 {g}_{\beta\beta} =  \Delta V   
 \frac{d}{2} 
 \frac{\rho}{m\beta^2},\nonumber \\   
 {g}_{u_i u_j}     
 =  \Delta V \rho\beta\delta_{ij}.    
 \label{gideal3}    
 \ea  
 Therefore, each volume element $\Delta V$ contributes  
 a squared distance $(dL)^2$ between two states 
 $(\rho,\beta,\vec u)$ and 
 $(\rho +d\rho,\beta +d\beta,\vec u + d\vec u)$ 
 given by  
 \be 
 \frac{\Delta V\rho}{m}  
 \left[ \left(\frac{d\rho}{\rho}\right)^2 + 
 \frac{d}{2} 
 \left(\frac{d\beta}{\beta}\right)^2 +  
 m\beta\, d\vec u\cdot d\vec u \right]. 
 \label{idealm}   
 \ee  
 This expression agrees with our expectations: 
 1) The metric coefficients are independent of $\vec u$, as 
 demanded by Galilean invariance\footnote{The 
 Galilean invariance of the metric in the  
 general case (non-ideal gas) is proven in Section 6, 
 confer (\ref{Galil}).};   
 2) Since no prefered or external scales are available 
 and taking into account the Galilean invariance, 
 the differentials must always appear 
 in the form of $d\rho/\rho$ and $d\beta/\beta$; the gas 
 temperature determines the scale for measuring the kinetic 
 energy associated with $d\vec u$; 
 3) Cross-terms like $d\rho\,d\vec u$ and $d\beta\,d\vec u$ 
 cannot appear due to the rotational symmetry of the gas  
 and 
 4) The factor multiplying the square brackets is just the 
 number of particles in $\Delta V$.\\ 
 What could not have been guessed beforehand, is the 
 absence of the cross-term $d\rho\, d\beta$ and the precise 
 form of the coefficients.

 One of the simplest applications of a metric tensor 
 is the computation of the norm of a tangent vector, 
 as in (\ref{velo}). 
 Now we can do this in the configuration space of the 
 moving fluid, i.e., we can compute the norm of 
 the local rate of evolution of our sytem, 
 that we denote by $F$, as   
 \be 
 F ^2 :=     
 {g}_{ij}   
 {\dot{\theta}}^i {\dot{\theta}}^j,  
 \label{F}   
  \ee   
 where the dots indicate the  
 time derivatives of the corresponding quantities 
 as given by the Euler  equations. 
 In order to distinguish between this generalized velocity $F$ 
 and the standard velocities $\vec v_k$ and $\vec u$, 
 we shall call this quantity {\em the rapidity}. 
 The rapidity squared is a scalar with dimension  
 ${\rm [time]}^{-2}$. 
 From (\ref{lower}) we see that it     
 can also be written  as 
 \be 
 F^2  =  
 {\dot{\eta}}_i {\dot{\theta}}^i ,   
 \ee  
 i.e., it is the contraction between the velocities 
 of the intensive variables with the velocities of 
 their corresponding conjugate extensive variables.

 Let us compute the rapidity associated with the material 
 derivatives in the Euler equations with no external 
 forcing, i.e., 
 \ba 
 \frac{d\rho}{dt} = -\rho \mbox{\,div\,} \vec u, \nonumber\\
 \rho c_v\frac{dT}{dt} = -P \mbox{\,div\,}\vec u, \nonumber\\ 
 \rho \frac{d\vec u}{dt} = 
 -\vec\nabla P , 
 \label{material} 
 \ea 
 where $c_v$ is the specific heat at constant volume. 
 In the case of an ideal gas, i.e., 
 making use of (\ref{gideal3}) and inserting (\ref{material}) 
 into (\ref{F}) leads to 
 \be 
 F^2= \frac{\Delta V\rho}{m} 
 \left[   
 \left( {1 + \frac{d}{2}\left(\frac{R'}{c_v}\right)^2}\right) 
 (\mbox{\,div\,}\vec u)^2  
 + R'T \left|{\vec \nabla \ln(\rho T)}\right|^2 
 \right] ,  
 \ee 
 where $R':=k/m$ is the specific constant of the gas. 
 \section{Flatness of the manifold} 
 \vspace*{-0.1cm} 
 Another standard application of a metric is the determination 
 not only of an infinitesimal distance, as in (\ref{idealm}), 
 but also of finite distances, say $L({\bf 1},{\bf 2})$ 
 between  positions 
 ${\bf 1} := (\rho_1,-\beta_1,\vec u_1)$ and 
 ${\bf 2} := (\rho_2,-\beta_2,\vec u_2)$. 
 This distance 
 is defined as the length of the shortest path connecting 
 ${\bf 1}$ with ${\bf 2}$, 
 \be   
 L({\bf 1},{\bf 2})
  = \min 
 \int_1^2{\! d\tau \, 
 \sqrt{   
 {g}_{ij}(\theta)   
 \,\frac{d\theta^i}{d\tau}  
 \frac{d\theta^j}{d\tau} }},
 \label{L}   
 \ee   
 where the minimum is taken over all paths from 
 ${\bf 1}$ to ${\bf 2}$ and $\tau$ is any parametrization 
 of these paths.\\ 
 This calculation is not a simple one when 
 there is no coordinate system in which the metric tensor 
 $g_{ij}(\theta)$ takes a simple form; this happens in the 
 case of our moving fluid since, as we will see in this 
 Section, (\ref{idealChrist}) and comments thereafter, 
 (\ref{gideal3}) is {\em not} flat.  
 Luckily, it turns out that the geometry of statistical 
 manifolds offers more interesting and useful possibilities 
 than (\ref{L});
 this is particularly so in the case of a probability 
 density of the form (\ref{probability}) as is our case. 
 In order to realize this, we need some concepts from 
 differential geometry; for the sake of completeness,
 these concepts are listed in Appendix B.

 Two simple calculations indicate that the connexions 
 described in Appendix B 
 may play an important role in the case of our moving fluid:\\   
 Let us compute first the $[ij;k]$ connexion compatible with 
 the Fisher metric (\ref{coordina}); from (\ref{LC}) one has 
 that, in the coordinate system 
 $(\theta^1,\dots ,\theta^{d+2}) = 
 (\gamma,-\beta,\vec\kappa)$,    
 \be 
 [ij;k] = \frac{1}{2} 
 \frac{\partial^3\ln {\cal Z}(\theta)}%
 {\partial \theta^i\partial\theta^j\partial \theta^k}. 
 \label{idealChrist} 
 \ee 
 One can compute the corresponding curvature tensor and 
 check that, even for the ideal gas metric 
 (\ref{gideal3})-(\ref{idealm}), 
 this connexion is {\em not} 
 flat.\\  
 Next, let us compute the skewness tensor $T_{ijk}$ generated by 
 (\ref{probability}), confer (\ref{third}); 
 one finds that 
 \be 
 T_{ijk}(\theta) =
 \frac{\partial^3\ln {\cal Z}(\theta)}%
 {\partial \theta^i\partial\theta^j\partial \theta^k}. 
 \label{expoT} 
 \ee  
 From these two last results, 
 we see that by taking $[ij;k] - (1/2)T_{ijk}$ one gets a 
 connexion that vanishes identically in this coordinate 
 system, i.e., {\em this connexion is flat}.\\ 
 It should be noted that the existence of a flat connexion 
 implies that there is a prefered family of coordinate 
 systems, in our case, the system of the intensive variables 
 $\theta$ and their linear combinations. The physical 
 relevance of these variables becomes even more evident 
 when one considers the phenomenon of phase-coexistence, 
 e.g., liquid-vapour coexistence: coexisting phases 
 correspond to flat portions of the thermodynamic potential 
 surface only in terms of the intensive variables. 

 Similarly, one can check that also $[ij;k] + (1/2)T_{ijk}$, 
 notice the 
 change in sign, vanishes identically in the {\em extensive} 
 variables coordinate system $\eta$,
  i.e., also this connexion is flat.   
 Connexions of the form $[ij;k] - (\alpha/2)T_{ijk}$ play an 
 important role in the theory of statistical manifolds; 
 they were introduced by Chentsov \cite{chentsov}, 
 Efron \cite{Efron} and Amari
 \cite[Chapter 3]{amari}.  
 In the third reference it is shown that if 
 such a connexion is flat for a certain $\alpha$ then it is also 
 flat for $-\alpha$. 
 In fact, the coordinate systems 
 $\theta = 
 (\gamma,-\beta,\vec\kappa)$ 
 and 
 $\eta = 
 (\langle N\rangle ,\langle H\rangle ,\langle\vec M\rangle )$ 
 play totally symmetric roles: 
 it can be shown \cite[Theorems 3.4 and 3.5]{amari} 
 that $g^{ij}$, the inverse 
 of the metric tensor, is also given by the second-order 
 derivatives of a function $\Phi(\eta)$, this time with respect to 
 the extensive variables $\eta$, i.e., 
 \be 
 g^{ij}(\eta) = 
  \frac{\partial^2\Phi(\eta)}%
 {\partial \eta_i\partial\eta_j},       
 \label{ginverse} 
 \ee       
 that the function $\Phi(\eta)$ 
 is nothing else but the Legendre transform 
 of $\ln{\cal Z}(\theta)$ and that the coordinate systems are  
 related to each other as in a Legendre transformation, i.e.,   
 \ba 
 \theta^i (\eta) = \frac{\partial\Phi}{\partial\eta_i} , 
 \label{theta}\\ 
 \Psi(\theta) + \Phi(\eta) - \theta^i\cdot\eta_i =0,
 \label{Legendre}  
 \ea 
 where we have introduced $\Psi := \ln {\cal Z}$,
 confer also (\ref{eta}). 
 Recall that all functions are assumed to be sufficiently 
 smooth so that (\ref{Legendre}) is good enough for the 
 definition of the Legendre transform; for more complicated 
 situations, see \cite{Fenchel}.

 We have shown then that the probability density of the moving 
 fluid 
 generates not only a metric tensor (\ref{coordina}) but also two 
 flat 
 connexions, namely $[ij;k] \pm (1/2)T_{ijk}$ and that all this is 
 closely related to a Legendre transformation of thermodynamic 
 potentials and variables. 
 Some interesting consequences of these facts are presented in the next 
 Section.  
 \section{Distance-like function} 
 \vspace*{-0.1cm} 
 While the computation of the distance $L$ in (\ref{L}) 
 implies difficult calculations, 
 it was pointed out by Amari \cite[Section 3.5]{amari} that, 
 when the connexions  
 $[ij;k] \pm (1/2)T_{ijk}$ 
 are flat, 
 it is possible to define a 
 distance-like function $D(\bf 1,\bf 2)$ 
 between positions 
 ${\bf 1} := (\gamma_1,-\beta_1,\vec \kappa_1)$ and 
 ${\bf 2} := (\gamma_2,-\beta_2,\vec \kappa_2)$ 
 as follows 
 \be 
  D({\bf 1},{\bf 2}) := 
 \Psi ({\bf 1}) + \Phi({\bf 2}) - 
 \theta^i({\bf 1})\cdot\eta_i({\bf 2}), 
 \label{amariD}  
 \ee 
 with 
 $\Psi := \ln {\cal Z}$ as in 
 (\ref{Legendre}), confer also (\ref{ginverse}), (\ref{theta}) 
 and (\ref{eta}). 

 Amari has shown that 
 this function shares some essential properties with    
 the usual distance functions, to wit\footnote{%
 The first property is a special instance of Fenchel's theorem 
 \cite{Fenchel}.}:\\  
 \ba  
 1)\qquad  
 D({\bf 1},{\bf 2}) \geq 0, \nonumber\\ 
 {\rm the\ equality\ holds\ when,\   
 and\ only\ when,}  
 \quad {\bf 1} ={\bf 2},
 \label{p1}\\     
 2)\qquad  
 D(\theta,\theta + d\theta) =  
 \nonumber\\ 
 \frac{1}{2}\,  
  g_{ij}(\theta)\, d\theta^i\, d\theta^j +  
 \frac{1}{2}\left\{ { 
 [i,j;k] + \frac{1}{ 3!} T_{ijk}(\theta) 
 } \right\} 
 \, d\theta^i\, d\theta^j\,  d\theta^k  
 + O(d\theta^4),\label{again} \\  
 3) \qquad D({\bf 1},{\bf 2})  =  
 D({\bf 1},Q) + D(Q,{\bf 2}),\nonumber\\  
 {\rm where\ }Q{\rm \ is\ connected\ to\ }{\bf 1}{\rm\ by\ a\ }
  \theta{\rm -geodesic\ and\   
 to\  }{\bf 2}{\rm\ by\ an\ } 
  \nonumber\\ 
  \eta{\rm  -geodesic\  
 and\ these\ two\ geodesics\    
  intersect\  orthogonally\ at\ }Q
  \label{path}\\  
 {\rm and\ } \qquad 4) \qquad {\rm The\  }  
 \min_{Q \in \Omega} D({\bf 1},Q)\nonumber   
 \\
 {\rm is\ obtained\ at\ a\ point\ on\ the\ boundary\ of\ 
 the\ smooth\ closed\ region\ }\Omega
 \nonumber\\ 
 {\rm\ that\ is\ the\  
 projection\ of\ }{\bf 1}{\rm\  along\ a\  }  
 \theta{\rm -geodesic\ orthogonal}\nonumber\\
 {\rm to\  the\   
 boundary\ of\ } \Omega. 
 \ea 
 The $\theta$-geodesics above is a linear 
 interpolation from $\theta({\bf 1})$ to $\theta(Q)$ 
 while the $\eta$-geodesic is a linear interpolation 
 from $\eta (Q)$ to $\eta({\bf 2})$. 
 The third property is a generalization   
 of Pythagoras' theorem to a space with two biorthogonal 
 coordinate bases, related to $\theta$ and $\eta$ in our case.  
 Analogously, the fourth property 
 generalizes the notion of projection to such a space. 
 The reader  should  
 refer to \cite[Section 3.5]{amari} for the 
 proofs of the properties listed above. \\  
 There is one important difference between the usual 
 distance functions and the $D$ defined above: 
 In general, the $D$ function is   
 {\em not} symmetric, i.e., $D({\bf 1},{\bf 2})\ne 
 D({\bf 2},{\bf 1})$.  
 This asymmetry is associated with the fact that the path 
 going from ${\bf 1}$ {\em first} along an $\eta$-geodesic 
 to $Q'$  and {\em then} along an orthogonal 
 $\theta$-geodesic to ${\bf 2}$ is, in general, different from the 
 path described in (\ref{path}). If one is set on defining a
 symmetric 
 distance, then taking the minimum of the values 
 $D({\bf 1},{\bf 2})$ and 
 $D({\bf 2},{\bf 1})$ seems to be the most satisfactory solution.\\ 
 From properties (\ref{p1}) and (\ref{again}), we recognize that 
 this $D$ may belong to the family of contrast functionals 
 discussed in Section 3. 
 In fact, 
 using (\ref{eta}), (\ref{coordina}), (\ref{theta}) and 
 (\ref{Legendre}), 
 one finds that
 $D({\bf 1},{\bf 2})$ in (\ref{amariD}) 
 can be written as    
 \be  
 D({\bf 1},{\bf 2}) 
 =(\beta_2-\beta_1) \langle H\rangle_1 -  
 (\vec\kappa_2-\vec\kappa_1)\cdot\langle\vec M\rangle_1 - 
 (\gamma_2-\gamma_1)\langle N\rangle_1 +  
 \Delta V (\beta_2 P_2 -\beta_1 P_1),   
 \ee  
 where $\langle\cdots\rangle_1$ indicates that an average is taken 
 over $p(N,E,\vec M;\theta_1)$. 
 One can check then that, in our case, the last expression for 
 $D({\bf 1}, {\bf 2})$ 
 is identical to  (\ref{D}) with 
 $f(z)= z \ln z$. 
 The circle is now complete.

 The Galilean invariance of $D({\bf 1},{\bf 2})$ 
 follows from introducing 
 (\ref{epsilon}), (\ref{kappa}) and (\ref{gamma}) 
 into the last expression above; one finds then that 
 the contrast functional (\ref{D}) or (\ref{amariD}) 
 can be written as 
 \be 
 D({\bf 1},{\bf 2}) =  
 (\beta_2-\beta_1) \langle N\rangle_1 \epsilon_1 +  
 \frac{m}{2} \beta_2 |\vec u_2 - \vec u_1|^2  
 \langle N\rangle_1 -   
 (\bar{\gamma}_2-\bar{\gamma}_1)\langle N\rangle_1 +  
 \Delta V (\beta_2 P_2 -\beta_1 P_1). 
 \label{Galil}  
 \ee 
 Combining this with (\ref{taylor}) and with (\ref{again}), 
 one sees  that  both $g_{ij}(\theta)$ and 
 $T_{ijk}(\theta)$ are always 
 independent of $\vec u$, i.e., Galilean invariant.\\ 
 Notice also that when the variations in the 
 thermodynamic variables can be ignored, 
 as is the case 
 in an (effectively) incompressible, isentropic flow, 
 the contrast reduces to 
 $ D({\bf 1},{\bf 2}) =
 (\Delta V\rho\beta/2) |\vec u_1 - \vec u_2|^2$ 
 with $\rho =\rho_1 =\rho_2$ and $\beta =\beta_1 = 
 \beta_2$,
 i.e., the kinetic-energy norm  
 times a conformal factor.\\  
 In the case of an ideal gas, the 
 contrast function reads:  
 \be 
 D({\bf 1},{\bf 2}) 
 =\frac{\rho_1}{m} \Delta V 
 \left[  
 \frac{m}{2} \beta_2 |\vec u_2 - \vec u_1|^2 +  
 \left( { \ln {\frac{\rho_1}{\rho_2}}  +  
 \frac{\rho_2}{\rho_1} -1 } \right) +  
 \frac{d}{2} \left( { \ln \frac{\beta_1}{\beta_2} +  
 \frac{\beta_2}{\beta_1} -1 }\right)  
  \right]. 
 \label{idealD} 
 \ee 

 \section{Related work}
 Before closing, we review and 
 comment some papers  
 that deal with related questions:\\ 
 Weinhold \cite{Weinhold} proposed 
 to use the matrix of second-order derivatives of the internal 
 energy as a metric in the space tangent to the 
 equation-of-state surface at an equilibrium point. 
 Some researchers tried then to attach a 
 physical meaning to the curvature tensor derived from the 
 Weinhold's metric-compatible connexion,  
 confer Section 5. This led to a lengthy 
 discussion which has been summarized in \cite{Andresen}; 
 all these researchers ignored the flat connexions 
 $[ij;k] \pm\frac{1}{2} T_{ijk}$ and the associated 
 distance-like function $D$ as discussed in Section 6.  
 In the context of the present article, Weinhold's metric 
 comes close to the Fisher metric (\ref{coordina}) while 
 Gilmore's approach, see \cite{Andresen}, is closer to 
 the distance-like function $D$ (\ref{amariD}).\\ 
 The Fisher metric was 
 introduced into thermodynamics by Ingarden  
 \cite{Ingarden1,Ingarden2}.   
 Janyszek and Mruga{\l}a \cite{Janyszek} studied the 
 curvature tensors of the corresponding metric-compatible 
 connexions and   
 tried to associate this curvature with  
 some physical properties; they did not consider 
 the flat connexions 
 $[ij;k] \pm\frac{1}{2} T_{ijk}$ neither the associated 
 distance-like function $D$.    
 
 On the 18th of December 1995, 
 I presented this paper at the Technical 
 University of Berlin. Prof.\ U.\ Simon informed me then 
 about affine and projective differential geometry 
 which deals, among other things, with conjugate 
 connexions like $[ij;k] \pm\frac{1}{2} T_{ijk}$ and their 
 flatness. Excellent, up-to-date reviews of this 
 branch of differential geometry are \cite{Simon} and 
 \cite{Nomizu}. 
 
 There exists a rich literature on the  
 Hamiltonian (symplectic) structure of 
 hydrodynamics that can also be 
 seen as one branch of differential geometry; 
 see, e.g., the review articles
 \cite{Salmon,Shepherd}, the references therein 
 and \cite{Zeitlin,Rouhi} for another interesting 
 application. 
 This approach is based on the Poisson bracket  
 and it does {\em not} require a 
 metric tensor. By contraposition, 
 the present article does not use the Poisson bracket 
 and leads to a metric tensor, flat connexions, etc;
 i.e., the two 
 approaches are not necessarily related and can be seen 
 as complementary. 
 In Arnold's analysis of incompressible flows \cite{Arnold}, 
 a metric plays an important role, however, 
 this metric is just the  metric 
 of the ambient $d$-dimensional space that appears in 
 the kinetic energy term, 
 i.e., is not the type of metric developed in the 
 present article. 
 There is  
 one instance of a Hamiltonian structure for 
 fluid dynamics that apparently 
 contains a candidate for a metric tensor 
 in the configuration space of a moving fluid; 
 this instance is discussed in Appendix C.

 \section{Summary and discussion} 
 \vspace*{-0.1cm} 
 In this article, it has been shown that the grand canonical 
 partition function (\ref{probability}) which describes the 
 thermal fluctuations of a moving fluid in local thermal equilibrium 
 generates a natural metric tensor $g_{ij}(\theta)$, two flat 
 connexions $[ij;k]\pm \frac{1}{2}T_{ijk}$ and 
 the corresponding distance-like 
 contrast function $D$, confer (\ref{coordina}), 
 (\ref{idealChrist}), (\ref{expoT}) and (\ref{amariD}).  
 In the case of an ideal gas, the explicit 
 expressions for  
 these quantities have been given in (\ref{idealm}) and in 
 (\ref{idealD}). 
 It was also shown that these quantities are Galilean invariant, 
 as they should, and that they reduce to the 
 kinetic-energy norm times a 
 conformal factor when one can ignore variations in 
 variables other than the velocity field, confer (\ref{Galil}). 
 Noteworthy aspects of the geometric structure are that the flat 
 connexions are {\em not} metric-compatible and that the 
 distance-like function $D$ is {\em not} symmetric. 
 As pointed out by one of the referees, the analysis presented 
 in this article can be applied to relativistic fluids, to 
 plasmas, etc. 
 In fact, some of the results listed above, e.g., 
 the particular form (\ref{coordina}) of the metric tensor, 
 as well as (\ref{ginverse})--(\ref{Legendre})  
 and the flatness of the connexions 
 $[ij;k]\pm \frac{1}{2}T_{ijk}$, are valid for all systems 
 characterized by Gibbs' probability distributions like 
 (\ref{pd}).

 The following considerations 
 are relevant with regard to further applications of our approach.  
 We have seen that all the components of the geometric structure 
 are generated from $\ln {\cal Z}=\Delta V\beta P$ and 
 since, in the absence of external forces, 
 the equation of state $P=P(\bar{\gamma},\beta)$ 
 is all that is needed in order to write down the 
 Euler equations of fluid motion, it is then clear   
 that the metric, connexions and contrast function 
 are closely related to the dynamical equations.
 When the system is far from thermal equilibrium, however, 
 a different approach may be more appropriate. For example, 
 when considering a fluid in a turbulent stationary state, 
 the corresponding probability density should be used. 
 Similarly, when studying the predictability properties of 
 systems in the presence of noise, a time-dependent probability 
 density should be employed \cite{yo}. 
 In this sense, the geometric structures we have described 
 are not universal;    
 the specific physical conditions 
 of the system under consideration determine whether a 
 thermal-equilibrium, a stationary-state or a time-dependent 
 probability density is the most appropriate starting point.

 \bc {\bf Acknowledgements} \ec 
 I would like to thank Robert Mureau for numerous conversations 
 and useful comments,  
 friends and colleagues, 
 in particular Gregory Falkovich and 
 Itamar Procaccia, 
 for their stimulating interest and Uriel Frisch for
 some practical advices.   
 This article is dedicated to Marta and to Siggy.   
 \appendix 
 \bc {\bf Appendix A: On Fisher's metric} \ec 
 \vspace*{-0.1cm} 
 Besides the clear meaning due to (\ref{taylor}), 
 the Fisher metric tensor (\ref{g}) 
 plays an important role in statistics   
 due to the following theorem \cite{Rao,Cramer}:\\  
 Suppose that we perform $n$ independent measurements of the  
 random variables $(N,E,\vec M)$ and use them in order  
 to estimate the value of the parameters $(\gamma,-\beta,\vec  
 \kappa)$ in (\ref{probability}). 
 Denote by $\{\hat\theta^i\}$ an unbiased 
 estimate of these parameters. Then, for large enough $n$, 
 the covariance of 
 these estimated values with respect to the exact values  
 $\{\theta^i\}=(\gamma,-\beta,\vec \kappa)$  
 has the following lower bound  
 \be  
 {\rm cov\  } \left[
 (\hat\theta^i -\theta^i) (\hat\theta^j -\theta^j) 
 \right] \geq \frac{1}{n} g^{ij}, 
 \ee  
 where $(g^{ij})$ is the inverse of the matrix $(g_{ij})$. 
 In other words, given a fixed tolerance error, 
 the larger the distance between two probability distributions 
 as measured by $g_{ij}$, 
 the smaller the number of measurements needed in order
 to distinguish between them. 
 
  Finally, one should notice the following identity,
 \be      
 g_{ij}(\theta)=  
 \left\langle {\partd {\ln p}{\theta^i}  
 \partd{\ln p}{\theta^j} }\right\rangle=      
 -\left\langle {
 \frac{\partial^2 \ln p}{\partial\theta^i\partial\theta^j} } 
 \right\rangle .      
 \label{apeng}      
 \ee  
 This identity is obtained as follows:  
 $\int\! dx\, p(x,\theta)  =1$ implies that 
 $\langle \partial
 \ln p(x,\theta)/\partial\theta^i\rangle =
 \int\! dx\,\partial p(x,\theta)/\partial\theta^i = 0$;     
 taking then the derivative of 
 $\langle \partial
 \ln p(x,\theta)/\partial\theta^i\rangle$ with respect to $\theta^j$
 leads to the identity in (\ref{apeng}). 
 (It is assumed that the order of derivation with respect to 
 the $\{\theta^i\}$ and integration over $x$ can be 
 interchanged.)
 \appendix 
 \bc {\bf Appendix B: On affine connexions} \ec
 In some cases, the geometry of a manifold is not 
 completely defined only by its
 metric tensor but also by another geometric object, 
 called the connexion of the manifold\footnote{When 
 this happens, we speak of a non-Riemannian 
 manifold.}, see, e.g., 
 \cite[Chapter 4, Sections 28 and 29]{geometry}
 and \cite[Chapters I--IX]{Erwin}. 
 Briefly stated, the connexion expresses mathematically  
 what it means ``to move a vector along a curve in such 
 a way that the vector remains constant", 
 i.e., it defines the parallel transport of a vector.\\  
 We list some important properties of the connexions that we 
 need in Section 5:\\ 
 $\bullet$\quad 
 Under a general transformation of the coordinates, a  
 connexion does {\em not} transform like a tensor, i.e., 
 a connexion is {\em not} a tensor. In particular: a 
 symmetric  
 connexion may vanish identically in a particular coordinate 
 system {\em without} vanishing in a different one while a 
 tensor that vanishes identically in a particular system of 
 coordinates does so in all coordinate systems.\\ 
 $\bullet$\quad 
 The result of adding a third-rank tensor to a connexion 
 is a connexion.\\  
 $\bullet$\quad
 We say that a manifold is {\em flat} with respect to a 
 connexion when every vector that is
 parallel transported along every closed path returns 
 to its original condition; otherwise, one says that the 
 manifold is curved.\\  
 $\bullet$\quad 
 A manifold is flat with respect to a connexion if, 
 and only if, it is possible to find a 
 coordinate system in which this connexion vanishes identically.\\ 
 $\bullet$\quad 
 Whether a given connexion is flat or not can be determined in a 
 coordinate-system independent way: a connexion is flat if, and only if, 
 it is symmetric 
 in its two first indices, i.e., torsionless, and 
 its curvature tensor vanishes identically.\\  
 $\bullet$\quad 
 The partial derivatives of a tensor are {\em not} tensors, 
 however, using the connexion it is possible to define the 
 so-called covariant derivatives which are tensors. 
 If the connexion vanishes identically in a given coordinate 
 system then, in that coordinate system, the covariant derivatives 
 coincide with the partial derivatives.\\ 
 $\bullet$\quad 
 There is only one connexion symmetric in its two first indices 
 such that the covariant 
 derivatives of the metric tensor vanish identically. 
 If $g_{ij}$ 
 denotes the metric tensor, then this connexion\footnote{%
 Actually, this connexion is obtained by raising the $k$-index 
 of the $[ij;k]$ in (\ref{LC}), i.e., by $g^{lk}[ij;k]$.} is given by 
 \be 
 \frac{1}{2} \left( \partd{g_{ik}}{\theta^j} 
 + \partd{g_{jk}}{\theta^i} - \partd{g_{ij}}{\theta^k} 
 \right). 
 \label{LC}  
 \ee 
 This is precisely the $[ij;k]$ appearing in the 
 third-order term in (\ref{taylor}), confer (\ref{christo}). 
 This connexion is called 
 the metric-compatible or Levi-Civita connexion. 
 For the proofs of these statements, refer to, e.g., 
 \cite[Chapter 4, Sections 28 and 29]{geometry}
 and \cite[Chapters I--IX]{Erwin}.
 \bc {\bf Appendix C: On the Dubrovin-Novikov 
 approach to flows} \ec 
 \vspace*{-0.1cm} 

 In an illuminating article \cite{DN1},  
 Dubrovin and Novikov introduced a novel  
 Hamiltonian formalism for  
 one-dimensional systems of hydrodynamic  
 type described by $s$ dynamical fields.   
 (An extensive review of related ideas and  
 developments can be found in \cite{DN3}.)    
 One notable aspect of this formalism is   
 the central role played by a symmetric   
 covariant tensor of type (2,0), call it   
 $\Lambda^{ij},\, 1\le  (i,j)\le s$:  
 when   
 $\det(\Lambda^{ij})\ne 0$, 
 the Jacobi identity 
 that the Poisson bracket has to satisfy implies that the   
 Levi-Civita connexion generated by $\Lambda_{ij}:= 
 (\Lambda^{-1})_{ij}$ 
 must be flat.  
 Moreover,  
 when  $\det(\Lambda^{ij})\ne 0$ 
 and other,   
 relatively mild conditions are satisfied,   
 it is possible  
 to reconstruct (up to a constant  
 factor) the tensor $\Lambda^{ij}$   
 directly  
 from the hydrodynamical equations of  
 motion \cite{Tsarev}.  
 Therefore, in such cases, it is  
 possible to associate with the  
 equations of motion, one metric in the   
 phase space of the fluid: the inverse of   
 $(\Lambda^{ij})$.   
 In practice, these symmetric tensors    
 cannot be used as metrics but in few,   
 unrealistic cases due to the following reasons:  
 1) When the fluid exists in a  
 $d$-dimensional space, $d$ different  
 symmetric tensors $\Lambda^{ij}$   
 must be introduced \cite{DN2} leading, in the  
 best case, to $d$ different metrics in  
 the fluid phase space;  
 2) Even in the case of a   
 one-dimensional physical space,   
 one often has that   
 $\det(\Lambda^{ij}) \equiv 0$,  
 this is the case, e.g., of a  
 one-dimensional nonbarotropic fluid  
 \cite{DN1,DN2},\cite[p. 59]{DN3}.   
 Moreover, it should be noticed that the  
 $\Lambda^{ij}$ tensor is part   
 of the Poisson bracket definition, therefore,   
 different dynamical systems generated  
 by different Hamiltonians, i.e., fluids with   
 different equations of state, but by the   
 same Poisson bracket share the same  
 $\Lambda^{ij}$;   
 this property is obviously not   
 desirable when one is trying to  
 identify the geometric structures  
 that characterize and distinguish  
 between different hydrodynamical  
 systems.  
 Similarly, some hydrodynamical  
 equations have a multi-Hamiltonian  
 formulation, i.e., the same set of  
 equations can be generated from  
 different Hamiltonians by means of  
 different Poisson brackets \cite{magri,Nutku};   
 in such cases,   
 each of these Poisson  
 brackets would lead to a different  
 metric tensor for the same   
 dynamical equations.   
  
\end{document}